\documentclass[11pt]{article}
\usepackage{amssymb}
\usepackage{epsfig}

\newcommand{\BE}{\begin{equation}}
\newcommand{\EE}{\end{equation}}
\begin{document}
\begin{titlepage}

\vspace*{1mm}
\begin{center}

   {\LARGE{\bf An effective vacuum refractive index from gravity and the present
    ether-drift experiments}}

\vspace*{14mm}
{\Large  M. Consoli and E. Costanzo}
\vspace*{4mm}\\
{\large
Istituto Nazionale di Fisica Nucleare, Sezione di Catania \\
Dipartimento di Fisica e Astronomia dell' Universit\`a di Catania \\
Via Santa Sofia 64, 95123 Catania, Italy \\ }
\end{center}
\begin{center}
{\bf Abstract}
\end{center}

Re-analyzing the data published by the Berlin and D\"usseldorf
ether-drift experiments, we have found a clean non-zero daily
average for the amplitude of the signal. The two experimental
values, ${A}_0\sim (10.5 \pm 1.3)\cdot 10^{-16}$ and $A_0\sim
(12.1\pm 2.2)\cdot 10^{-16}$ respectively, are entirely consistent
with the theoretical prediction $(9.7\pm 3.5)\cdot10^{-16}$ that is
obtained once the Robertson-Mansouri-Sexl anisotropy parameter is
expressed in terms of ${\cal N}_{\rm vacuum}$, the effective vacuum
refractive index that one would get, for an apparatus placed on the
Earth's surface, in a flat-space picture of gravity .
\end{titlepage}

\section{Introduction}

The present generation of ether-drift experiments, combining the
possibility of active rotations of the apparatus with the use of
cryogenic optical resonators, is currently pushing the relative
accuracy of the measured frequency shifts to the level ${\cal
O}(10^{-16})$. As we shall try to illustrate, this level of accuracy
could be crucial to determine basic properties of the vacuum such as
its space-time structure.

To this end, we'll present a re-analysis of the observations
reported in Refs.\cite{peters,schiller} for the anisotropy of the
speed of light in the vacuum. This re-analysis leads to two
conclusions: i) both experiments exhibit a non-zero daily average
for the amplitude of the signal ii) the magnitude of this average
amplitude is entirely consistent with the theoretical Robertson -
Mansouri - Sexl (RMS) \cite{robertson,mansouri} anisotropy parameter
\BE \label{first} |(1/2-\beta+\delta)|_{\rm th}\sim 3({\cal N}_{\rm
vacuum}-1)\sim 42\cdot 10^{-10}\EE  that one would get
\cite{pagano,pla,cimento} in terms of ${\cal N}_{\rm vacuum}$, the
effective vacuum refractive index that arises in a flat-space
picture of gravity.

The plane of the paper is as follows. In Sect.2, we shall first
illustrate the basic formalism and report the experimental data of
Refs. \cite{peters,schiller}. Then, in Sect.3, we shall use these
data to deduce the daily average amplitude of the signal for the two
experiments. Further, in Sect.4 we shall compare these experimental
values with the theoretical prediction that one would get, if there
is a preferred frame, in a flat-space description of gravity.
Finally, in Sect.5, we shall present our summary and conclusions.

\section{Basic formalism and experimental data}

The experimental data reported in Ref.\cite{peters} refer to 15
short-period observations, performed from December 2004 to April
2005, while the observations of Ref.\cite{schiller} refer to a
single short-period observation, taken around February 8th 2005. The
starting point for our analysis is the expression for the relative
frequency shift of two optical resonators at a given time $t$. For
the Berlin experiment \cite{peters}, this can be expressed as
\BE \label{basic2}
      {{\Delta \nu (t)}\over{\nu_0}} = {S}(t)\sin 2\omega_{\rm rot}t +
      {C}(t)\cos 2\omega_{\rm rot}t
\EE
where $\omega_{\rm rot}$ is the rotation frequency of one resonator
with respect to the other which is kept fixed in the laboratory and
oriented north-south. The Fourier expansions of $S(t)$ and $C(t)$
are predicted to be  \BE \label{amorse1}
      {S}(t) = S_0 +
      {S}_{s1}\sin\tau +{S}_{c1} \cos\tau
       + {S}_{s2}\sin(2\tau) +{S}_{c2} \cos(2\tau)
\EE
\BE
\label{amorse2}
      {C}(t) = {C}_0 +
      {C}_{s1}\sin\tau +{C}_{c1} \cos\tau
       + {C}_{s2}\sin(2\tau) +{C}_{c2} \cos(2\tau)
\EE
where $\tau=\omega_{\rm sid}t$ is the sidereal time of the
observation in degrees and $\omega_{\rm sid}\sim
{{2\pi}\over{23^{h}56'}}$. Introducing the colatitude of the
laboratory $\chi$, and the unknown average velocity, right ascension
and declination of the cosmic motion with respect to a hypothetical
preferred frame (respectively $V$, $\alpha$ and $\gamma$), one finds
the expressions reported in Table I of Ref.~\cite{peters} \BE
\label{C0}
      {C}_0 =-
      {{K \sin^2\chi}\over{8}} (3 \cos 2{\gamma} -1)
\EE
\BE
\label{CS1}
      {C}_{s1}= {{1}\over{4}}K
      \sin 2{\gamma} \sin{\alpha} \sin 2\chi
      ~~~~~~~~~~~~~~~~~~~~~~~~
     {C}_{c1}={{1}\over{4}}K \sin 2{\gamma}
      \cos{\alpha} \sin 2\chi
\EE
\BE
\label{CS2}
      {C}_{s2} = {{1}\over{4}}K \cos^2{\gamma}
      \sin2{\alpha}  (1+ \cos^2\chi)
~~~~~~~~~~~~
      {C}_{c2} = {{1}\over{4}}K \cos^2{\gamma}
      \cos2{\alpha} (1+ \cos^2\chi)
\EE
where \BE \label{cappa}
 K=(1/2-\beta+\delta){{V^2}\over{c^2}}
\EE and $(1/2-\beta+\delta)$  indicates the RMS
\cite{robertson,mansouri} anisotropy parameter. The corresponding
$S-$quantities are also given by ($S_0=0$) \BE
\label{s1}{S}_{s1}=-{{ {C}_{c1} } \over {\cos\chi}}~~~~~~~
{S}_{c1}={ {{C}_{s1} }\over{ \cos\chi\ }} \EE \BE \label{s2}
{S}_{s2}= -{{2\cos\chi}\over{1+\cos^2\chi}}{C}_{c2}~~~~~~~{S}_{c2} = {{2\cos\chi}\over{1+\cos^2\chi}}{C}_{s2}\EE For the D\"usseldorf
experiment of Ref.\cite{schiller}, one should just re-nominate the
two sets \BE(C_0,C_{s1},C_{c1},C_{s2},C_{c2})\to
(C_0,C_1,C_2,C_3,C_4)\EE
 \BE (S_0,S_{s1},S_{c1},S_{s2},S_{c2})\to
(B_0,B_1,B_2,B_3,B_4)\EE and introduce an overall factor of two for
the frequency shift since, in this case, two orthogonal cavities are
maintained in a state of active rotation.

As suggested by the same authors, it is safer to concentrate on the
observed time modulation of the signal, i.e. on the quantities
${C}_{s1},{C}_{c1},{C}_{s2},{C}_{c2}$ and on their
${S}$-counterparts. In fact, the constant components ${C}_0 $ and
$S_0=B_0$ are likely affected by spurious systematic effects such as
thermal drift. The experimental C-coefficients are reported in Table
1 for Ref.\cite{schiller} and in Table 2 for Ref.\cite{peters}
(these latter numerical values have been extracted from  Fig.3
Ref.\cite{peters}).
\begin{table*}
\caption{The experimental $C-$coefficients as reported in
Ref.\cite{schiller}. }
\begin{center}
\begin{tabular}{clll}
\hline\hline $C_{s1}[{\rm x}10^{-16}]$ & $C_{c1}[{\rm x}10^{-16}]$ &
$C_{s2}[{\rm x}10^{-16}]$ &
$C_{c2}[{\rm x}10^{-16}]$   \\
\hline
$-3.0\pm 2.0$ &  $11.0\pm 2.5$ &  $1.0\pm2.5$ &  $0.1\pm 2.5$ \\
\hline\hline
\end{tabular}
\end{center}
\end{table*}
\begin{table*}
\caption{The experimental $C-$coefficients as extracted from Fig.3
of Ref.\cite{peters}. }
\begin{center}
\begin{tabular}{clll}
\hline\hline $C_{s1}[{\rm x}10^{-16}]$ & $C_{c1}[{\rm x}10^{-16}]$ &
$C_{s2}[{\rm x}10^{-16}]$ &
$C_{c2}[{\rm x}10^{-16}]$   \\
\hline
$-2.7\pm4.5$ &  $5.3\pm 4.8$ &  $-3.2\pm4.7$ &  $1.2\pm 4.2$ \\
$-18.6\pm6.5$ &  $8.9\pm 6.4$ &  $-11.4\pm6.5$ &  $-5.0\pm 6.4$ \\
$-0.7\pm3.9$ &  $5.3\pm 3.6$ &  $5.0\pm3.5$ &  $1.6\pm 3.8$ \\
$6.1\pm4.6$ &  $0.0\pm 4.8$ &  $-8.1\pm4.8$ &  $-4.0\pm 4.6$ \\
$2.0\pm8.6$ &  $1.3\pm 7.7$ &  $16.1\pm8.0$ &  $-3.3\pm 7.2$ \\
$3.0\pm5.8$ &  $4.6\pm 5.9$ &  $8.6\pm5.9$ &  $-6.9\pm 5.9$ \\
$0.0\pm5.4$ &  $-9.5\pm 5.7$ &  $-5.5\pm5.6$ &  $-3.5\pm 5.4$ \\
$-1.1\pm8.1$ &  $11.0\pm 7.9$ &  $0.9\pm8.3$ &  $18.6\pm 7.9$ \\
$8.6\pm6.5$ &  $2.7\pm 6.7$ &  $4.3\pm6.5$ &  $-12.4\pm 6.4$ \\
$-4.8\pm4.8$ &  $-5.1\pm 4.8$ &  $3.8\pm4.7$ &  $-5.2\pm 4.7$ \\
$5.7\pm3.2$ &  $3.0\pm 3.4$ &  $-6.3\pm3.2$ &  $0.0\pm 3.5$ \\
$4.8\pm8.0$ &  $0.0\pm 7.0$ &  $0.0\pm7.6$ &  $1.5\pm 7.7$ \\
$3.0\pm4.3$ &  $-5.9\pm 4.3$ &  $-2.1\pm4.4$ &  $14.1\pm 4.3$ \\
$-4.5\pm4.4$ &  $-2.3\pm 4.5$ &  $4.1\pm4.3$ &  $3.2\pm 4.3$ \\
$0.0\pm3.6$ &  $4.6\pm 3.4$ &  $0.6\pm3.2$ &  $4.9\pm 3.3$ \\
\hline\hline
\end{tabular}
\end{center}
\end{table*}
\vfill\eject The relevant numbers for the S-coefficients of
Ref.\cite{peters} are reported in our Table 3. The S-coefficients of
Ref.\cite{schiller} were constrained, in the fits to the data, to
their theoretical predictions in Eqs.(\ref{s1}) and (\ref{s2}). Thus
their values will be deduced from Table 1 using these relations.


\begin{table*}
\caption{The experimental $S-$coefficients as extracted from Fig. 3
of Ref.\cite{peters}.}
\begin{center}
\begin{tabular}{clll}
\hline\hline $S_{s1}[{\rm x}10^{-16}]$ & $S_{c1}[{\rm x}10^{-16}]$ &
$S_{s2}[{\rm x}10^{-16}]$ &
$S_{c2}[{\rm x}10^{-16}]$   \\
\hline
$11.2\pm4.7$ &  $11.9\pm 4.9$ &  $1.8\pm4.9$ &  $0.8\pm 4.5$ \\
$1.8\pm6.5$ &  $-4.3\pm 6.5$ &  $6.4\pm6.4$ &  $1.8\pm 6.4$ \\
$-3.3\pm3.8$ &  $2.9\pm 3.8$ &  $-5.9\pm3.8$ &  $4.6\pm 4.0$ \\
$12.7\pm5.1$ &  $14.3\pm 5.5$ &  $-1.9\pm5.3$ &  $-3.3\pm 5.1$ \\
$4.7\pm8.4$ &  $-6.9\pm 7.3$ &  $-1.8\pm8.0$ &  $-7.8\pm 7.0$ \\
$5.2\pm5.8$ &  $-3.0\pm 5.9$ &  $7.1\pm5.9$ &  $-5.9\pm 5.8$ \\
$11.1\pm5.3$ &  $-13.4\pm 5.4$ &  $-4.5\pm5.5$ &  $-9.8\pm 5.5$ \\
$-12.1\pm8.9$ &  $0.0\pm 8.8$ &  $-3.1\pm9.0$ &  $1.4\pm 8.9$ \\
$-4.8\pm6.3$ &  $6.5\pm 6.4$ &  $-8.1\pm6.3$ &  $3.5\pm 6.5$ \\
$9.8\pm5.0$ &  $4.8\pm 5.0$ &  $1.9\pm5.0$ &  $-9.2\pm 4.8$ \\
$0.0\pm3.2$ &  $-3.9\pm 3.6$ &  $1.0\pm3.1$ &  $-2.2\pm 3.4$ \\
$-12.7\pm7.7$ &  $8.5\pm 6.8$ &  $-8.3\pm7.2$ &  $-7.1\pm 7.4$ \\
$-7.9\pm4.7$ &  $-4.3\pm 4.8$ &  $-1.9\pm4.8$ &  $-6.2\pm 4.7$ \\
$16.1\pm4.9$ &  $12.0\pm 5.2$ &  $2.9\pm4.9$ &  $-9.6\pm 4.8$ \\
$13.9\pm3.9$ &  $-7.0\pm 3.4$ &  $-3.3\pm3.5$ &  $3.0\pm 3.6$ \\
\hline\hline
\end{tabular}
\end{center}
\end{table*}

\section{The daily average amplitude of the signal}

For our analysis, we shall re-write Eq.(\ref{basic2}) as follows \BE
\label{basic3}
      {{\Delta \nu (t)}\over{\nu_0}} = A(t)\cos (2\omega_{\rm rot}t -2\theta_0(t))
\EE with \BE \label{interms}
C(t)=A(t)\cos2\theta_0(t)~~~~~~~~S(t)=A(t)\sin2\theta_0(t)\EE
$\theta_0(t)$ representing the instantaneous direction of a
hypothetical ether-drift effect in the plane of the interferometer.

Within the RMS model, the amplitude of the signal (a
positive-definite quantity) can be expressed in terms of $v(t)$, the
magnitude of the projection of the cosmic Earth's velocity in the
plane of the interferometer as
\BE \label{amplitude1}
       A(t)= {{1}\over{2}}|(1/2 -\beta +\delta)| {{v^2(t) }\over{c^2}} ,
\EE
To compute $v(t)$, we shall use the expressions given by Nassau and
Morse \cite{nassau}.  These are valid for short-period observations,
as those performed in Refs.\cite{peters,schiller}, where the
kinematical parameters of the cosmic velocity ${\bf{V}}$ are not
appreciably modified by the Earth's orbital motion around the Sun.
In this case, by introducing the latitude of the laboratory $\phi$,
the right ascension $\alpha$ and the declination $\gamma$ associated
to ${\bf{V}}$, the magnitude of the Earth's velocity in the plane of
the interferometer is defined by the two equations \cite{nassau}
\BE
       \cos z(t)= \sin\gamma\sin \phi + \cos\gamma
       \cos\phi \cos(\tau-\alpha)
\EE
and
\BE \label{vearth}
       v(t)=V \sin z(t) ,
\EE
$z=z(t)$ being the zenithal distance of ${\bf{V}}$.

 Replacing Eq.~(\ref{vearth}) into Eq.~(\ref{amplitude1}) and
adopting a notation of the type in
Eqs.(\ref{amorse1})-(\ref{amorse2}), we obtain
\BE \label{amorse}
       A(t) = A_0 +
       A_1\sin\tau +A_2 \cos\tau
        +  A_3\sin(2\tau) +A_4 \cos(2\tau)
\EE
where ($\chi=90^o-\phi$)
\BE \label{aa0}
       A_0 ={{1}\over{2}} |K|
       \left(1- \sin^2\gamma\cos^2\chi
       - {{1}\over{2}} \cos^2\gamma\sin^2\chi \right)
\EE
\BE \label{a1}
       A_1=-{{1}\over{4}}|K| \sin 2\gamma
       \sin\alpha \sin 2\chi
~~~~~~~~~~~~~~~
       A_2=-{{1}\over{4}}|K| \sin 2\gamma
       \cos\alpha \sin 2\chi
\EE
\BE \label{a3}
       A_3=-{{1}\over{4}} |K| \cos^2 \gamma
       \sin 2\alpha \sin^2 \chi
~~~~~~~~~~~~~~~
       A_4=-{{1}\over{4}} |K| \cos^2 \gamma
       \cos 2\alpha \sin^2 \chi
\EE
Since $A_0$ was not explicitly given by the authors of
Ref.\cite{peters,schiller}, we shall now deduce its value from their
published data that indeed have been obtained with experimental
sessions extending over integer multiples of 24 hours in length
\cite{peters}. The daily averaging of the signal (here denoted by
$\langle..\rangle$), when used in Eq.(\ref{amorse}) produces the
relation \BE \label{amplitude0}\langle A^2(t) \rangle= A^2_0+
{{1}\over{2}}(A^2_{1}+A^2_{2}+A^2_{3}+A^2_{4})\EE On the other hand,
using Eqs.(\ref{amorse1}), (\ref{amorse2}) and (\ref{interms}), one
also obtains
 \BE
\label{amplitude}\langle A^2(t) \rangle= C^2_0 + S^2_0 +
{{1}\over{2}}(C^2_{11}+S^2_{11}+C^2_{22}+S^2_{22})\EE where we have
introduced the combinations
\BE \label{csid}
      {C}_{11}\equiv \sqrt{{C}^2_{s1}
      + {C}^2_{c1}}
~~~~~~~~~~~~~~~~
      {C}_{22}\equiv \sqrt{{C}^2_{s2}
      + {C}^2_{c2}}
\EE
 \BE \label{s2sid}
      {S}_{11}\equiv \sqrt{{S}^2_{s1}
      + {S}^2_{c1}}
~~~~~~~~~~~~~~~~
 {S}_{22}\equiv \sqrt{{S}^2_{s2}
      + {S}^2_{c2}}
\EE
As one can check, replacing the expressions (\ref{aa0})-(\ref{a3}),
Eq.(\ref{amplitude0}) gives exactly the same result that one would
obtain replacing the values for the C- and S- coefficients in
Eq.(\ref{amplitude}). Therefore, one can combine the two relations
and get \BE \label{final} A^2_0(1+r)= C^2_0 + S^2_0 +
{{1}\over{2}}(C^2_{11}+S^2_{11}+C^2_{22}+S^2_{22}) \EE with \BE
r\equiv {{1}\over{2A^2_0}}(A^2_{1}+A^2_{2}+A^2_{3}+A^2_{4}) \EE

To evaluate $A_0$ we shall proceed as follows. On the one hand, we
shall compute the ratio $r=r(\gamma,\chi)$ using the theoretical
expressions Eqs.(\ref{aa0})-(\ref{a3}). This gives \BE \label{range}
0\leq r\leq 0.40 \EE for the latitude of the two laboratories in the
full range $0 \leq |\gamma|\leq \pi/2$. On the other hand, we shall
adopt  the point of view of the authors of
Refs.\cite{peters,schiller} that, even when large non-zero values of
$C_0$ and $S_0$ are obtained (compare with the value $C_0=(-59.0 \pm
3.4 \pm 3.0)\cdot 10^{-16}$ of Ref.\cite{schiller} and with the
large scatter of the data reported in Fig.3 of Ref.\cite{peters}),
tend to consider these individual determinations as spurious
effects. This means to set in Eq.(\ref{final}) \BE S_0 =\langle
A(t)\sin 2\theta_0(t)\rangle\sim 0\EE  \BE C_0 =\langle A(t)\cos
2\theta_0(t)\rangle\sim 0\EE The resulting average daily amplitude,
determined in terms of $C_{11}$, $S_{11}$, $C_{22}$ and $ S_{22}$
alone, provides, in any case, a lower bound to its true experimental
value. The data for the various coefficients are reported in our
Tables 4 and 5 together with the quantity \BE \label{Q} Q= \sqrt{
{{1}\over{2}}(C^2_{11}+S^2_{11}+C^2_{22}+S^2_{22})} \sim
A_0\sqrt{1+r} \EE from which, taking into account the numerical
range of $r$ in Eq.(\ref{range}), we finally get \BE \label{AQ}
A_0\sim (0.92 \pm 0.08)Q \EE
\begin{table*}
\caption{ The experimental values of Ref.\cite{schiller} for the
combinations of $C-$ and $S-$ coefficients defined in
Eqs.(\ref{csid})-(\ref{s2sid}) and the resulting $Q$ from
Eq.(\ref{Q}). For simplicity, we report symmetrical errors. The
values for the S-coefficients, constrained in the fits to the data
to their theoretical predictions in Eqs.(\ref{s1}) and (\ref{s2}),
have been deduced from Table 1 using these relations.}
\begin{center}
\begin{tabular}{cllll}
\hline\hline $ {C}_{11}  [{\rm x}10^{-16}]$ & $ {C}_{22}  [{\rm
x}10^{-16}]$ & $ {S}_{11}  [{\rm x}10^{-16}]$ &
$ {S}_{22}  [{\rm x}10^{-16}]$& $ Q [{\rm x}10^{-16}]$ \\
\hline
$11.4\pm 2.5 $ &  $1.0\pm 2.5$ &  $14.7\pm3.2$ &  $1.0\pm 2.5$ &${13.2}\pm 2.1$\\

\hline\hline
\end{tabular}
\end{center}
\end{table*}
\begin{table*}
\caption{ The experimental values of Ref.\cite{peters} for the
combinations of $C-$ and $S-$ coefficients defined in
Eqs.(\ref{csid})-(\ref{s2sid}) and the resulting  $Q$ from
Eq.(\ref{Q}). For simplicity, we report symmetrical errors.}
\begin{center}
\begin{tabular}{cllll}
\hline\hline $ {C}_{11}  [{\rm x}10^{-16}]$ & $ {C}_{22}  [{\rm
x}10^{-16}]$ & $ {S}_{11}  [{\rm x}10^{-16}]$ &
$ {S}_{22}  [{\rm x}10^{-16}]$ & $ Q [{\rm x}10^{-16}]$  \\
\hline
$5.9\pm4.7$ &  $3.5\pm 4.6$ &  $16.3\pm4.8$ &  $2.0\pm 4.9$& ${12.6}\pm 3.5$ \\
$20.6\pm6.4$ &  $12.5\pm 6.5$ &  $4.6\pm6.5$ &  $6.6\pm 6.4$& ${17.8} \pm 4.7$ \\
$5.3\pm3.6$ &  $5.3\pm 3.6$ &  $4.4\pm3.8$ &  $7.5\pm 3.8$ & ${8.1}\pm 2.8$ \\
$6.1\pm4.6$ &  $9.0\pm 4.8$ &  $19.1\pm5.3$ &  $3.8\pm 5.1$ &${15.7}\pm 3.8$\\
$2.4\pm8.4$ &  $16.5\pm 8.0$ &  $8.4\pm7.7$ &  $8.0\pm 7.1$ &${14.2}\pm 6.1$\\
$5.5\pm5.9$ &  $11.0\pm 5.9$ &  $6.0\pm5.9$ &  $9.2\pm 5.9$ &${11.6}\pm 4.5 $\\
$9.5\pm5.7$ &  $6.5\pm 5.5$ &  $17.4\pm5.4$ &  $10.7\pm 5.5$ &${16.6} \pm 4.0 $\\
$11.0\pm7.9$ &  $18.7\pm 7.9$ &  $12.1\pm8.9$ &  $3.4\pm 9.0$ &${17.7} \pm 6.2$\\
$9.1\pm6.5$ &  $13.1\pm 6.4$ &  $8.1\pm6.4$ &  $8.8\pm 6.4$ &${14.1}\pm 4.8$\\
$7.0\pm4.8$ &  $6.5\pm 4.7$ &  $10.9\pm5.0$ &  $9.4\pm 4.8$ &${12.2}\pm 3.7$\\
$6.4\pm3.1$ &  $6.3\pm 3.2$ &  $3.9\pm3.6$ &  $2.4\pm 3.4$ &${7.0}\pm 2.4$\\
$4.8\pm8.0$ &  $1.5\pm 7.7$ &  $15.3\pm7.4$ &  $10.9\pm 7.3$ &${13.7}\pm 5.8$\\
$6.6\pm4.3$ &  $14.3\pm 4.3$ &  $9.0\pm4.7$ &  $6.5\pm 4.7$ &${13.6}\pm 3.3 $\\
$5.1\pm4.5$ &  $5.2\pm 4.3$ &  $20.0\pm5.0$ &  $10.0\pm 4.8$ &${16.6} \pm 3.6$\\
$4.6\pm3.4$ &  $5.0\pm 3.3$ &  $15.6\pm3.8$ &  $4.4\pm 3.5$ &${12.4}\pm 2.7$\\
\hline\hline
\end{tabular}
\end{center}
\end{table*}
For a more precise determination of  $Q$ for the experiment of
Ref.\cite{peters}, we observe that the values reported in Table 5
exhibit a good degree of statistical consistency. \vfill\eject This
can be checked through the chi-square of the weighted averages over
the 15 observation periods \BE \label{av1} {C}_{11}= (6.7 \pm
1.2)\cdot 10^{-16}~~~~~~ {C}_{22}= (7.6 \pm 1.2)\cdot
10^{-16}~~~~~~~~~~~\EE \BE\label{av2} {S}_{11} = (11.0 \pm 1.3)\cdot
10^{-16}~~~~~~ { S}_{22}= (6.3 \pm 1.3)\cdot 10^{-16}~~~~~~~~~~\EE
which is always of order unity. Using Eqs.(\ref{Q}) and (\ref{AQ})
these values give an average  $A_0$ for the 15 observation periods
of Ref.\cite{peters} \BE \label{mean1} {A}_0\sim (10.5 \pm 1.3)\cdot
10^{-16}~~~~~~~~~~~~~~~ \EE in good agreement with the value \BE
\label{mean2} A_0\sim (12.1\pm 2.2)\cdot 10^{-16}~~~~~~~~~~~~~~~ \EE
of Ref.\cite{schiller}.

\section{An effective refractive index for the vacuum}

In this section, we shall point out that the two experimental values
in Eqs.(\ref{mean1}) and (\ref{mean2}) are well consistent with the
theoretical prediction \BE \label{theory1} A^{\rm th}_0 \sim
{{1}\over{2}} |1/2-\beta+\delta|_{\rm th} {{v^2}\over{c^2}} \sim
(9.7 \pm 3.5)\cdot 10^{-16} \EE of Refs.\cite{pla,cimento}. This was
obtained, in connection with the RMS parameter \cite{pagano}
$|1/2-\beta+\delta|_{\rm th}\sim 42\cdot 10^{-10}$, after inserting
the average cosmic velocity (projected in the plane of the
interferometer) $v=(204 \pm 36)$ km/s that derives from a
re-analysis \cite{pla,cimento} of the classical ether-drift
experiments. Due to this rather large theoretical uncertainty, the
different locations of the various laboratories and any other
kinematical property of the cosmic motion can be neglected in a
first approximation.

For a proper comparison, we also remind that in
Refs.\cite{pla,cimento}, the frequency shift was parameterized as
\BE {{\Delta\nu(\theta)}\over{\nu_0}}= |1/2-\beta+\delta|_{\rm th}
{{v^2}\over{c^2}} \cos 2\theta\EE This relation is appropriate for a
symmetrical apparatus with two rotating orthogonal lasers, as in the
D\"usseldorf experiment \cite{schiller}), and gives an average
amplitude \BE 2A_0 \sim (19 \pm 7)\cdot 10^{-16}\EE

The theoretical prediction for the RMS parameter was obtained
starting from the formal analogy that one can establish between
General Relativity and a flat-space description with re-defined
masses, space-time units and an effective vacuum refractive index.
This alternative approach, see for instance Wilson \cite{wilson},
Gordon \cite{gordon}, Rosen \cite{rosen}, Dicke \cite{dicke},
Puthoff \cite{puthoff} and even Einstein himself \cite{pre}, before
his formulation of a metric theory of gravity, in spite of the deep
conceptual differences, produces an equivalent description of the
phenomena in a weak gravitational field.

The substantial phenomenological equivalence of the two approaches
was well summarized by Atkinson as follows \cite{atkinson} : "It is
possible, on the one hand, to postulate that the velocity of light
is a universal constant, to define {\it natural} clocks and
measuring rods as the standards by which space and time are to be
judged and then to discover from measurement that space-time is {\it
really} non-Euclidean. Alternatively, one can {\it define} space as
Euclidean and time as the same everywhere, and discover (from
exactly the same measurements) how the velocity of light and natural
clocks, rods and particle inertias {\it really} behave in the
neighborhood of large masses."

This formal equivalence, which is preserved by the weak-field
classical tests, is interesting in itself and deserves to be
explored. In fact, "...it is not unreasonable to wonder whether it
may not be better to give up the geometric approach to gravitation
for the sake of obtaining a more uniform treatment for all the
various fields of force that are found in nature" \cite{rosen}.

For a quantitative test, one can start from the Equivalence
Principle \cite{pre}. According to it, for an observer placed in a
freely falling frame, local Lorentz invariance is valid. Therefore,
given two space-time events that differ by $(dx,dy,dz,dt)$, and the
space-time metric \BE ds^2=c^2dt^2- (dx^2+dy^2+dz^2)\EE one gets
from $ds^2=0$ the same speed of light that one would get in the
absence of any gravitational effect.

For an observer placed on the Earth's surface, for which the only
gravitational field with respect to which the laboratory is not in
free fall is that of the Earth, both General Relativity and the
flat-space approach predict the weak-field, isotropic form of the
metric \BE \label{iso} ds^2=c^2dt^2g_{44}-g_{11}(dx^2+dy^2+dz^2)=c^2d\tau^2-dl^2\EE where
$g_{44}=(1-{{2GM}\over{c^2R}})$, $g_{11}=(1+{{2GM}\over{c^2R}})$,
 $G$ being Newton's constant and $M$ and $R$ the Earth's mass and
radius. Here $d\tau$ and $dl$ denote respectively the elements of
"proper" time and "proper" length in terms of which, in General
Relativity, one would again deduce from $ds^2=0$ the same universal
value ${{dl}\over{d\tau}}=c$.

However, in the flat-space approach the condition $ds^2=0$ is
interpreted in terms of an effective refractive index for the vacuum
 \BE{\cal N}_{\rm vacuum}- 1 \sim {{2GM}\over{c^2R}}\sim 14\cdot
10^{-10}\EE as if Euclidean space would be filled by a very rarefied
medium. Is it possible to distinguish {\it experimentally} between
the two different interpretations ?

To this end, let us recall that a moving dielectric medium acts on
light as an effective gravitational field \cite{gordon,leonard} and
that, propagating in the "gravitational medium", light can be seen
isotropic by only one inertial frame \cite{volkov}, say $\Sigma$.
Thus the following question naturally arises: according to the
ether-drift experiments, does $\Sigma$ coincide with the Earth's
frame or with the hypothetical preferred frame of Lorentzian
relativity ? In the former case, corresponding to no anisotropy of
the two-way speed of light in the vacuum, the equivalence between
General Relativity and the gravitational-medium picture would
persist. In the latter case, using Lorentz transformations, one
predicts an anisotropy governed by the RMS parameter
\cite{pagano,pla,cimento} \BE \label{theory2}|1/2-\beta+\delta|_{\rm
th}\sim 3({\cal N}_{\rm vacuum}- 1 )\sim 42\cdot 10^{-10}\EE whose
observation would uniquely single out the flat-space scenario. More
precisely, one would be driven to conclude that the isotropic form
of the metric Eq.(\ref{iso}) does not hold for an observer placed on
the Earth's surface and applies to some other frame (whose physical
interpretation, within standard General Relativity, is not obvious).
For this reason, the present ether-drift experiments, with their
${\cal O}(10^{-16})$ accuracy, represent precision probes of the
vacuum and of its space-time structure.

\section{Summary and conclusions}
In this paper, we have presented a re-analysis of two ether-drift
experiments \cite{peters,schiller} that, employing rotating
cryogenic optical resonators, attempt to establish the isotropy of
the speed of light in the vacuum to a level of accuracy ${\cal
O}(10^{-16})$. For our re-analysis, we started by re-writing
Eq.(\ref{basic2}) as \BE
      {{\Delta \nu (t)}\over{\nu_0}} =  A(t)\cos (2\omega_{\rm rot}t -2\theta_0(t))
\EE and assuming, as the authors of Refs.\cite{peters,schiller},
that experimental results providing large non-zero values for either
$\langle C(t) \rangle= C_0$ or $\langle S(t)\rangle =S_0$ in
Eqs.(\ref{amorse1}) and (\ref{amorse2}) should be interpreted as
spurious effects (e.g. due to thermal drift, non-uniformity of the
rotating cavity speed, misalignment of the cavity rotation
axis,...).

With this assumption, the daily average for the amplitude of the
signal $A_0=\langle A(t)\rangle$ can be expressed as\BE A_0\sim
(0.92\pm 0.08)Q\EE where \BE Q= \sqrt{
{{1}\over{2}}(C^2_{11}+S^2_{11}+C^2_{22}+S^2_{22})} \EE is given in
terms of the coefficients $C_{11},C_{22}, S_{11}, S_{22}$ defined in
Eqs.(\ref{csid})-(\ref{s2sid}). They represent the simplest
rotationally invariant combinations one can form with the elementary
coefficients $C_{s1}, C_{c1}, C_{s2}, C_{c2}$ and with their
S-counterparts. As stressed by the authors of Refs.
\cite{peters,schiller}, these coefficients, that reflect the time
modulation of the signal, should be much less affected by spurious
effects than $C_0$ and $S_0$. Therefore, computing $A_0$ in this way
should be completely safe. In any case, comparing with the full
result in Eq.(\ref{final}), our method provides a lower bound for
the true experimental value of $A_0$.

Now, the two resulting experimental determinations in
Eqs.(\ref{mean1}) and (\ref{mean2}), namely ${A}_0\sim (10.5 \pm
1.3)\cdot 10^{-16}$ and $A_0\sim (12.1\pm 2.2)\cdot 10^{-16}$  are
in good agreement with each other and with the theoretical
prediction $(9.7\pm 3.5)\cdot 10^{-16} $ of
Refs.\cite{pagano,pla,cimento} that is obtained, in a flat-space
description of gravity, in the presence of a preferred reference
frame. As far as we can see, this non-trivial level of consistency
means that a non-zero anisotropy of the speed of light in the vacuum
has actually been measured in these experiments with values of the
RMS anisotropy parameter that are one order of magnitude larger than
the presently quoted ones.

For instance, in Ref.\cite{peters} the set ($V\sim$370 km/s,
$\alpha\sim 168^o$, $\gamma\sim -6^o$), corresponding to parameters
obtained from a dipole fit to the COBE data, was assumed from the
very beginning in the analysis of the data. In this case, fixing
$V\sim$ 370 km/s and replacing the value of the RMS parameter from
Ref.\cite{peters} $|(1/2-\beta+\delta)|\sim (2\pm 2)\cdot 10^{-10}$
in Eq.(\ref{cappa}), one would expect $|K|\sim (3\pm 3)\cdot
10^{-16}$ and $C_{11}=(0.15\pm 0.15)\cdot10^{-16}$, $S_{11}=(0.20\pm
0.20)\cdot10^{-16}$, $C_{22}=(1.2\pm1.2)\cdot10^{-16}$,
$S_{22}=(1.2\pm1.2)\cdot10^{-16}$.

These {\it expectations} should be compared with the actual
experimental values reported in Table 5 and with their weighted
averages  \BE {C}_{11}= (6.7 \pm 1.2)\cdot 10^{-16}~~~~~~ {C}_{22}(7.6 \pm 1.2)\cdot 10^{-16}~~~~~~~~~~~\EE \BE {S}_{11} = (11.0 \pm
1.3)\cdot 10^{-16}~~~~~~ { S}_{22}= (6.3 \pm 1.3)\cdot
10^{-16}~~~~~~~~~~\EE For this reason, in our opinion, the very
small RMS parameter of Ref.\cite{peters} (and of
Ref.\cite{schiller}) rather than reflecting the smallness of the
signal, originates from accidental cancellations among the various
entries. These might be due to several reasons. For instance, to a
wrong input choice for the kinematical parameters
$(V,\alpha,\gamma)$ used in the fits or to the procedure used to fix
the relative phases for the various parameter pairs (see note [13]
of Ref.\cite{peters}). These phases are essential to obtain
consistent values for the right ascension $\alpha$ and the sign of
$\gamma$. In any case, even a substantial level of phase error among
different experimental sessions, that can produce vanishing
inter-session averages for $C_{s1}, C_{c1}, C_{s2}, C_{c2}$ and
their S-counterparts, will not affect the rotationally invariant
combinations $C_{11}, C_{22}, S_{11}, S_{22}$ and our determination
of $A_0$.

To conclude, motivated by the fundamental nature of the questions
concerning the vacuum and its space-time structure, we have
undertaken a careful re-analysis of the data that leads to the
observed values of $A_0$ in Eqs.(\ref{mean1}) and (\ref{mean2}).
Since these results are entirely consistent with the theoretical
prediction Eq.(\ref{theory1}), we are driven to conclude that the
data support both the existence of a preferred frame and a
flat-space description of gravity. At the same time, the novelty of
this conclusion emphasizes the importance of comparing different
approaches and points of view to achieve a full understanding of the
underlying physical problem. \vfill\eject

\end{document}